\documentclass[prd,aps,secnumarabic,superscriptaddress,floatfix,preprintnumbers]{revtex4}
\usepackage[dvips]{graphicx,color}
\usepackage{dcolumn}
\def\sc{\mbox{\rule[-5pt]{0pt}{16pt}}}

\def\sb{\mbox{\rule{0pt}{11pt}}}

\def\sw{\mbox{\rule{24pt}{0pt}}}
\def\al{\alpha}

\def\Up{\Upsilon}

\begin{document}
\preprint{NSF-KITP-12-001} \vspace{12pt}
\title{Potential model results for the newly discovered $\chi_b(3P)$ states}
\author{Stanley F. Radford}\email{sradford@brockport.edu}
\affiliation{Department of Physics, The College at Brockport, State
University of New York, Brockport, NY 14420 \\ and \\
Kavli Institute for Theoretical Physics, Santa Barbara, CA 93106}
\email{sradford@brockport.edu}
\author{Wayne W. Repko}
\affiliation{Department of Physics and Astronomy, Michigan State
University, East Lansing, MI 48824}\email{repko@pa.msu.edu}
\date{\today}
\begin{abstract} The ATLAS Collaboration has recently announced the
discovery of the 3P states of the $\Up$ system, $\chi_b(3P)$, with
mass of 10.539$\pm$0.004(stat.)$\pm$0.008(syst.) GeV. In a previous
investigation of the $\Up$ system, in the context of a comprehensive
one-loop potential model, we calculated the masses of these states,
but did not include them.  We present those results, and others for
the n = 3 states here.
\end{abstract}
\maketitle
\section{INTRODUCTION}
A recent measurement by the ATLAS Collaboration \cite{atlas} has
located the spin-weighted mass barycenter of the n = 3 P states of
the $b\bar{b}$ system, the $\chi_b(3P)$, at a mass of
10.539$\pm$0.004(stat.)$\pm$0.008(syst.) GeV.  This mass value is
somewhat larger than previous model predictions \cite {kr, mz}.

In a recent paper, \cite{rr1}, we showed that the perturbative
treatment of a model consisting of a relativistic kinetic energy
term, a linear confining term including its scalar relativistic
corrections and the complete unsoftened perturbative one-loop
quantum chromodynamic short distance potential was able to reproduce
the overall spectrum of the $b\bar{b}$ system, including the
measured hyperfine structure of the S states, as well as its
radiative decays, with good accuracy.

Since the point of that paper was to recalculate the hyperfine
splittings, we did not include results for the 3P states. With the
ATLAS discovery of these states, it seems worthwhile to present our
theoretical results for the individual 3P states and their
splittings.
\section{RESULTS}
The parameters we use are shown in Table \ref{params} \cite{rr1}.
\begin{table}[h]
\centering
\begin{tabular}{ldddd}
\toprule
  &\multicolumn{1}{r}{\sb Model parameters}\sw\\
\hline
\sc$A$ (GeV$^2$) & 0.175 \\
\hline
\sc $\al_S$      & 0.295 \\
\hline
\sc $m_q$ (GeV)  & 5.33  \\
\hline
\sc $\mu$ (GeV)  & 4.82  \\
\hline
\sc $f_V$        & 0.00  \\
\botrule
\end{tabular}
\caption{Fitted parameters in the  $b\bar{b}$
potential.\label{params}}
\end{table}
It should be noted that in Table \ref{params}, $\al_S$ is the strong
coupling constant in the GR scheme \cite{gr1} and $f_V$ is the
fraction of vector coupling arising in the relativistic corrections.
The other parameters have their usual meanings. See Ref.
\cite{rr1,rr2} for a complete discussion of the potential model we
use.

\begin{table}[ht]\centering
\begin{tabular}{lddd} \toprule
\multicolumn{1}{c}{\sc}$m_{b\bar{b}}$\,(MeV)    & \multicolumn{1}{c}{ Our}& \multicolumn{1}{c}{ Expt } \\
\hline
\sb$\eta_b(3S)$\mbox{\rule{12pt}{0pt}}   & 10333.9  &  \\
\hline
\sb$\Up(3S)$         & 10364.2  &   10355.2\pm 0.5  \\
\hline
\sb$\chi_{b\,0}(3P)$ & 10516.0  &   \\
\hline
\sb$\chi_{b\,1}(3P)$ & 10538.1  &     \\
\hline
\sb$\chi_{b\,2}(3P)$ & 10552.9  &   \\
\hline
\sb$h_b(3P)$           & 10544.4  &  \\
\hline
\sb$3^3D_1$     & 10693.5     &  \\
\hline
\sb$3^3D_2$       & 10700.4    &   \\
\hline
\sb$3^3D_3$       & 10705.9    &    \\
\hline
\sb$3^1D_2$       & 10701.7    &    \\
\hline
\botrule
\end{tabular}
\caption{Results for the n=3 states of the $b\bar{b}$ spectrum.
Experimental data from \protect\cite{pdg}.}\label{bottomspec}
\end{table}

\section{DISCUSSION AND CONCLUSION}

Although the individual 3P spin states are not resolved in the Atlas
measurements, they are able to locate the spin-weighted average at
10.539$\pm$0.004(stat.)$\pm$0.008(syst.) GeV.  From the results in
Table \ref{bottomspec} we find the spin-weighted average of the the
triplet P states to be 10.5439 GeV, in good agreement with the ATLAS
data.  In addition, we calculate the intra-multiplet splittings as:
$\chi_{b\,2}(3P)$-$\chi_{b\,1}(3P)$=14.8 Mev;
$\chi_{b\,1}(3P)$-$\chi_{b\,0}(3P)$=22.1 MeV.  Previous model
predictions for the center of mass are somewhat lower, at 10.5197
MeV \cite{kr} and 10.525 MeV \cite{mz}, but, overall, potential
model predictions provide a reasonable description of the $\Up$
states below the $B\bar{B}$ threshold. We anticipate that further
data will provide information on the spin splittings of these
states.

\begin{acknowledgments}
SFR thanks the Kavli Institute for Theoretical Physics for its
hospitality during June 2009 and July 2010. This research was
supported in part by the National Science Foundation under Grants
PHY-1068020 and PHY-0551164.
\end{acknowledgments}


\begin{thebibliography}{99}

\bibitem{atlas} The ATLAS Collaboration, arXiv:1112.5154v1[hep-ex].
\bibitem{kr} W. Kwong and J. L. Rosner, Phys Rev D {\bf 38}, 279
(1988).
\bibitem{mz} L. Moytka and K. Zaleski, Eur. Phys. J. C {\bf 4}, 107
(1998).
\bibitem{rr1} S. F. Radford and W. W. Repko, Nucl. Phys. A {\bf 865},69 (2011).
\bibitem{rr2} S. F. Radford and W. W. Repko, Phys. Rev. D {\bf 75}, 074031 (2007).
\bibitem{gr1} S. N. Gupta and S. F. Radford, Phys. Rev. D {\bf25},
2690 (1982).
\bibitem{pdg} C. Amsler {\it et al.} (Particle Data Group), Phys. Lett. {\bf B667}, 1
(2008) (URL:http://pdg.lbl.gov).
\end{thebibliography}
\end{document}